\documentclass[reprint,amsmath,amssymb,aps]{revtex4-1}
\usepackage{graphicx}
\usepackage{subfig}
\usepackage{caption}
\usepackage{dcolumn}% Align table columns on decimal point
\usepackage{bm}% bold math
\usepackage{array}
 
\newcommand\B{\rule[-2.0ex]{0pt}{0pt}}
\newcommand\T{\rule{0pt}{6.2ex}}
\newcommand\TS{\rule{0pt}{4ex}} 
\newcolumntype{M}[1]{>{\centering\arraybackslash}m{#1}}
\newcolumntype{C}{ >{\centering\arraybackslash} m{2cm} }
\newcolumntype{D}{ >{\centering\arraybackslash} m{1cm} }
\newcommand{\specialcell}[2][c]{%
  \begin{tabular}[#1]{@{}c@{}}#2\end{tabular}}
\usepackage{hyperref}
\hypersetup{
	colorlinks=true, %set true if you want colored links
	linktoc=all,     %set to all if you want both sections and subsections linked
	linkcolor=blue,  %choose some color if you want links to stand out
}

\newcommand{\gp}{\gamma_\phi}
\newcommand{\op}{\Omega_\phi}

\begin{document}

\title{Dynamical Systems Analysis of K-essence Model}
%\thanks{A footnote to the article title}

\author{Abhijit Chakraborty}
\email{abhijit.iiserk@gmail.com}
\author{Anandamohan Ghosh}
\email{anandamohan@iiserkol.ac.in}
\author{Narayan Banerjee}%
\email{narayan@iiserkol.ac.in}
\affiliation{Indian Institute of Science Education and Research Kolkata\\
Mohanpur Campus, West Bengal 741246, India
}
\date{\today}% It is always \today, today,

\begin{abstract}
In the present work we investigate the stability of the k-essence models allowing upto quadratic terms of the kinetic energy. The system of field equations is written as an autonomous system in terms of dimensionless variables and the stability criteria of the equilibria have been extensively investigated. The results strongly indicate that cosmologically consistent models dynamically evolve towards the quintessence model, a stable solution with a canonical form of the dark energy.
\begin{description}
\item[PACS numbers] 98.80.-k; 95.36.+x
\end{description}
\end{abstract}

\maketitle

\section{\label{sec:1}Introduction}

A wide variety of exotic matter, called the dark energy, have been suggested over the last twenty years in order to account for the present accelerated expansion of the universe\cite{schmidt, riess, perl, knopp}. Introduction of a cosmological constant $\Lambda$, in spite of its simplicity, can indeed successfully take care of the role of the dark energy\cite{paddy}. However, the estimated value of $\Lambda$ is extremely small compared to the theoretically predicted value and new models continue to be proposed. Various scalar field models \cite{sami, varun, martin} are definitely the most popular and effective models. In these models a field called quintessence is incorporated such that a positive potential, which dominates over the kinetic part, can give rise to an effective negative pressure and drive the acceleration. 

An exotic form of a scalar field with a non-canonical kinetic sector, had been used to drive the early inflation\cite{picon, garriga}. The kinetic sector is taken as a function $f=f(\phi, X)$ where $\phi$ is the field and $X=\frac{1}{2}{\phi}^{,\mu}{\phi}_{,\mu}$ is the standard canonical kinetic energy. This kinetic sector can have a negative coefficient and play a key role in the violation of the energy condition. Raychaudhuri equation tells us that the energy condition, $\rho + 3p \geq 0$, where $\rho$ and $p$ are the total energy density and total effective pressure, has to be violated for an accelerated expansion of an isotropic distribution. In a quintessence field, the violation of energy density is engineered by a negative pressure. This kind of an exotic matter where the kinetic sector is given as $f(X,\phi)$, now called a k-essence, was employed to drive the late time acceleration as well\cite{chiba, picon1}. For a brief but comprehensive overview of the k-essence, we refer to the work of Armendariz-Picon, Mukhanov and Steinhardt\cite{picon2}. Scherrer showed the relevance of a {\it purely kinetic} k-essence in the unified dark sector\cite{scherrer}. Malquarti {\it et al} showed that a k-essence can hardly be distinguished observationally from a standard quintessence\cite{malq} but it might have some imprints on the perturbation spectrum. A reconstruction of a k-essence model had been attempted by Sen\cite{anjan} and by Matsumoto and Nojiri\cite{nojiri}. \\

The motivation of the present work is to look at the conditions for stability of a class of k-essence models. Although there is a lot of work on the stability of various scalar field models like quintessence\cite{luis, nandan1, anjan1, odin1} and its classes like tracker or freezing\cite{soma, nandan2}, chameleon fields\cite{nandan3, odin2} etc., not much of work on the stability of k-essence models are found in the literature. Abramo and Pinto-Neto\cite{pintoneto} discussed the conditions for stability of those k-essence models which can cross the phantom divide (the equation of state parameter attains a value less that -1). They showed that the phantom nature is not in fact supported by k-essence. Yang and Xiang-Ting reported that the stability of k-essence depends crucially on the potential\cite{yang}. The latter work is for a class of k-essence models where the potential is coupled with the kinetic part in the form $V(\phi) F(X)$ where $F(X)$ is a function of the kinetic energy. \\

The present work is an exhaustive study of the stability of the k-essence models where $f(X,\phi)$ contains upto  quadratic orders of $X$. The method of dynamical systems analysis is used for this purpose. An excellent review of the dynamical systems as applied to cosmology can be found in the recent work by Bahamonde {\it et al}\cite{bahamonde}. The present analysis very clearly indicates that those k-essence models are favored which eventually evolves into the canonical quintessence models. \\

The paper is organized as follows. Section II contains a description of the model and sets up the autonomous system. Section III and IV deal with the actual analysis of the dynamical systems, with two categorization of the coupling parameters of the theory. Section V contains the concluding remarks.

\section{\label{sec:SoE} The Model}
In quintessence, the late-time acceleration is driven by the potential energy of the scalar field. If the potential energy dominates over the kinetic energy so that the effective pressure is sufficiently negative, the deceleration parameter $q = - \frac{a\ddot{a}}{{\dot{a}}^2}$ can have a negative value. However, it is possible to have accelerated expansion by introducing non-canonical kinetic energy terms in the Lagrangian. Such models are known as k-essence. In this work, we consider a Lagrangian \eqref{K_lag} which is a polynomial of degree 2 in the kinetic energy $X = \frac{1}{2} \partial_\mu\phi\partial^\mu\phi$ with the coefficients as functions of the scalar field $\phi$, 
\begin{equation}\label{K_lag}
\mathcal{L}= \alpha(\phi)X + \beta(\phi)X^2 - V(\phi)
\end{equation}
where $V(\phi)$ is the potential. The coefficients $\alpha$ and $\beta$ are taken as functions of the scalar field in order to make them dynamic. The use of a  polynomial of degree 2 in kinetic energy inducing accelerated expansion is common in existing literature \cite{jorge,Guendelman}. The motivation of this work is to determine if there is any constraint that can be imposed on these coefficients via the stability analysis. \\

We consider a spatially flat FRW metric to describe the universe. As the spacetime is spatially homogeneous and isotropic, the scalar field is a function of the cosmic time alone. Then the kinetic energy $X$ becomes $X = \frac{1}{2}\dot{\phi}^2$. The Friedmann equations for this universe containing both scalar field and a perfect fluid are
\begin{subequations}\label{FEH}
\begin{align}
& H^2 = \frac{1}{3}(\rho_m + \rho_\phi), \label{FEH1}\\
&\dot{H} = - \frac{1}{2}(\rho_m + \rho_\phi + \textup{p}_m + \textup{p}_\phi), \label{FEH2}
\end{align}
\end{subequations}
where $H$ is the Hubble parameter. $\rho_m$ and $\textup{p}_m$ are the energy density and pressure of the dark matter distributions respectively.   $\rho_\phi$ and $\textup{p}_\phi$ are the contributions from the scalar field to the energy density and the pressure sectors, respectively and are given by
\begin{subequations}\label{pe_K}
	\begin{align}
	&\rho_\phi = \alpha(\phi)X + 3\beta(\phi)X^2 + V(\phi), \label{ed_k} \\
	&\textup{p}_\phi = \alpha(\phi)X + \beta(\phi)X^2 - V(\phi). \label{pre_k}
	\end{align}
\end{subequations}
Here we have used the units with $8\pi G=1$. The baryonic matter that fills the universe is considered to be dust with $\textup{p}_m = 0$. Taking variation of the action w.r.t. the scalar field $\phi$ we arrive at the Klein-Gordon equation assuming the form 
\begin{widetext}
\begin{equation}\label{KG}
\ddot{\phi} \big[\alpha(\phi)+3\beta(\phi)\dot{\phi}^2\big] + \alpha^\prime(\phi)\frac{\dot{\phi}^2}{2} + 3\beta^\prime(\phi)\frac{\dot{\phi}^4}{4} + 3H\dot{\phi}\big[\alpha(\phi)+\beta(\phi)\dot{\phi}^2\big] + V^\prime(\phi) =0.
\end{equation}
\end{widetext}

If the fluid sector and the scalar field do not interact and conserve separately, the previous equation is not an independent equation and can be derived using the Friedmann equations. 
%Here a prime denotes the derivative w.r.t. the scalar field. 
This system has five unknowns viz. $a$, $\phi$, $\alpha(\phi)$, $\beta(\phi)$, and $V(\phi)$. To write down the set of autonomous system of differential equations we define five dimensionless variables,
\begin{eqnarray}
x &=& \frac{\alpha(\phi)\dot{\phi}^2}{6H^2},
~~y = \frac{\beta(\phi)\dot{\phi}^4}{12H^2}, \nonumber\\
b &=& \frac{V(\phi)}{3H^2}, 
~~~\lambda = \frac{1}{\alpha}\frac{d\alpha}{d\phi}\frac{\dot{\phi}}{H}, 
~~\delta = \frac{1}{\beta}\frac{d\beta}{d\phi}\frac{\dot{\phi}}{H}. \nonumber
\end{eqnarray}
The variables can be related to the original system variables by identifying that $x$ corresponds to the term linear in kinetic energy while $y$ corresponds to the term quadratic in kinetic energy. $b$ is related to the potential energy term in the Lagrangian, $\lambda$ and $\delta$ are measures of the steepness of the coefficients $\alpha(\phi)$ and $\beta(\phi)$ respectively and indicate how they change in time.
From the Friedmann equations \eqref{FEH} and the K-G equation \eqref{KG} now written in terms of dimensionless variables
we obtain the following dynamical system
\begin{subequations}\label{diff_eq_k}
\begin{eqnarray}
		 x^\prime &=& 3x {\cal G}+\lambda x- 2x {\cal F}, \label{diff_eq_k1} \\
	 y^\prime &=& 3y {\cal G} +\delta y- 4y {\cal F}, \label{diff_eq_k2} \\
	 b^\prime &=& 3b {\cal G} + \sigma b, \label{diff_eq_k3}\\
	 \lambda^\prime &=& \frac{3}{2}\lambda {\cal G} -\lambda^2(1-\Gamma)-\lambda {\cal F}, \label{diff_eq_k4}\\
	 \delta^\prime &=& \frac{3}{2}\delta {\cal G} -\delta^2(1-\tau) - \delta {\cal F}, \label{diff_eq_k5}\\
	{\rm with~~}  {\cal F} &=& \Bigg[\frac{3(x+2y)}{x+6y}+\frac{\lambda x+3\delta y+\sigma b}{2(x+6y)},\Bigg] \nonumber \\
	{\rm and~~}  {\cal G} &=& (x+y-b+1). \nonumber %\label{diff_eq_k1}
\end{eqnarray}
\end{subequations}

Here $\Gamma = \frac{\alpha(\frac{d^2\alpha}{d\phi^2})}{(\frac{d\alpha}{d\phi})^2}$, $\tau = \frac{\beta(\frac{d^2\beta}{d\phi^2})}{(\frac{d\beta}{d\phi})^2}$, and $\sigma = \frac{d(\ln\hspace{1mm} V)}{dN}$ are parameters. A prime denotes a  derivative with respect to $N$ where $N=\ln(a/a_0)$,  $a_0$ being the current value of the scale factor.
For the dynamical system analysis of the above system of equations, we will  first consider $\lambda$ and $\delta$ as parameters and in the subsequent section consider them as dynamical variables.

\section{\label{case1} Fixed coupling parameters}
The coupling parameters, $\lambda$ and $\delta$, as defined in the previous section, are actually the fractional rate of growth of the coupling parameters $\alpha (\phi)$ and $\beta (\phi)$, scaled by the Hubble parameter $H$. If these are treated as constant parameters, the system \eqref{diff_eq_k} effectively reduces to a 3-dimensional system as the left hand sides of \eqref{diff_eq_k4} and \eqref{diff_eq_k5} are identically zero. \\

If $\lambda$ and $\delta$ are constants, one trivial solution is that both are zero. We would rather fix them as constants $\lambda_0$ and $\delta_0$ respectively which takes care of the trivial solution as a special case. However, for the non-null solutions, the constant values of $\lambda$ and $\delta$ are connected by \\

\begin{equation}\label{lam-del}
\lambda_0 (1-\Gamma) = \delta_0(1-\tau).	
\end{equation}

For their constant values, the defining equation for $\lambda$ and $\delta$ can be easily integrated and one can check that the constraint (\ref{lam-del}) is trivially satisfied. Also one can note that particular values of the parameters will impose constraints like $\lambda_0 = \frac{\frac{3}{2} G - F}{1-\Gamma}$ and $\delta_0 = \frac{\frac{3}{2} G - F}{1-\tau}$. But the quantities $G, F, \Gamma , \tau$ remain sufficiently arbitrary as four quantities are connected by two equations.

\subsection*{Pure k-essence: $V(\phi) = 0$}
If we consider the case with zero potential energy, i.e. only the kinetic energy terms contribute to the Lagrangian, equation \eqref{diff_eq_k3} is also identically zero with $b=0$. The fixed points and the corresponding eigenvalues of this 2-dimensional system are shown in \autoref{tab:tab1}.

\begin{table*}
	\caption{\textbf{Fixed points for the 2-dimensional system with zero potential}}\label{tab:tab1}
	\begin{tabular}{c|c|c|c|D|D|c}
		\hline Point &	x & y & Eigenvalues & $\Omega_\phi$ & $\gamma_\phi$ & Stability Condition\\ \hline
		\hline A. & 0 & $\frac{1}{3}$ & \{$\frac{1}{2}(4-\delta + 2\lambda$), 1\} \TS\B &1 &$\frac{4}{3}$ & unstable.\\
		\hline B. & 1 & 0 & \{3, ($-6+\delta-2\lambda$)\} \TS\B  & 1 & 1 & unstable.\\
		\hline C. & $\frac{\delta-2\lambda-4}{2}$ & $\frac{6-\delta+2\lambda}{6}$ \T &\specialcell{\{$\frac{(2\lambda-\delta+6)(2\lambda-\delta+4)}{-2\lambda+\delta-8}$, \\ $(\delta-2\lambda-3)$\}} \B & 1 & $\frac{\delta-2\lambda}{3}$ &$\delta<2\lambda+3.$\\\hline
	\end{tabular} 
\end{table*}

The system has at least one positive eigenvalue for both point A and point B leading to the fact that A and B cannot be stable fixed points. They are either unstable nodes or saddle points depending on the values of $\lambda$ and $\delta$. But the fixed point C can give rise to a stable node if the following condition is satisfied
\begin{equation}\label{1cond_stab}
2\lambda>\delta-3.
\end{equation}

\begin{figure}
\centering
\includegraphics[width=0.7\linewidth]{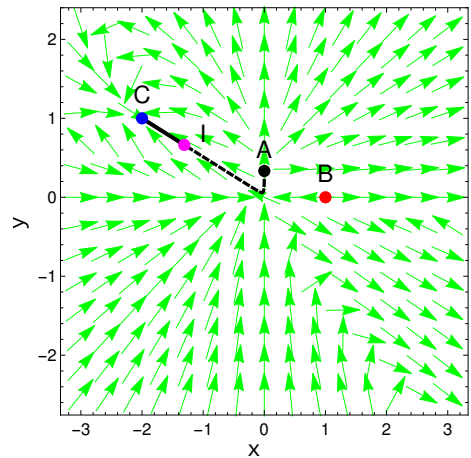}
\caption{Direction field plot for the 2D system with fixed points A,B and C from \autoref{tab:tab1} shown. I denotes the initial condition obtained from \autoref{init}. The solid and dashed black line shows the time evolution of the trajectory in future and past respectively.}
\label{fig:dir_field}
\end{figure}

Given \eqref{1cond_stab} is satisfied, the value of $x$ in case of the fixed point C becomes negative as $(\delta-2\lambda-4)$ becomes negative. However, $y$ remains positive because ($6-\delta+2\lambda) > 3$. From the definition of $x$, the only way for it to be negative is if the coefficient $\alpha(\phi)$ is negative. The direction field of the system suggests that a system that ends with a negative $x$ and positive $y$ has to start to with a negative $x$ and positive $y$ (\autoref{fig:dir_field}). This imposes a constraint on the nature of the coefficients $\alpha(\phi)$ and $\beta(\phi)$
\begin{equation}\label{alpbet_con}
\alpha(\phi) <0 ,\hspace {1cm} \beta(\phi)>0.
\end{equation}

With zero potential, the scalar field energy density parameter ($\Omega_\phi$) and effective equation of state (EoS) parameter ($\gamma_\phi$)  are given by,
\begin{subequations}\label{newvar_k}
	\begin{align}
	\Omega_\phi &= \frac{\rho_\phi}{3H^2} = x + 3y \label{newvar_k1}\\
	\gamma_\phi &= 1+w_\phi = 1+ \frac{\textup{p}_\phi}{\rho_\phi} = \frac{2x+4y}{x + 3y} \label{newvar_k2}
	\end{align}
\end{subequations}
where we have used equation \eqref{pre_k} and \eqref{ed_k} as the expression for the pressure and energy density with $V(\phi) = 0$. The values for $\Omega_\phi$ and $\gamma_\phi$ are taken as,
\begin{subequations}\label{init}
	\begin{align}
	\Omega_\phi &\approx 0.68, \label{init1}\\
	\gamma_\phi &\approx 0.05, \label{init2}
	\end{align}
\end{subequations} 
which are consistent with the current observations \cite{ade}.
Solving equation \eqref{newvar_k} with the values in \eqref{init1} and \eqref{init2} yields $x = -1.309$ and $y = 0.663$. The negative value of $x$ suggests that the universe, with its current scalar field energy density and EOS parameter will evolve to point C eventually and condition \eqref{1cond_stab} is satisfied to culminate in a non-diverging fate of the universe.

For ($\gamma_\phi < 0$), we have the equation of state $w_\phi <-1$ giving rise to a universe expanding faster than exponential, eventually leading up to the \textit{`The Big Rip'} \cite{caldwell}. This model of dark energy is known as the phantom field. However, if one wants to avoid `the big rip' and restrict our system with the condition $\gp\geq0$, we encounter another condition on the parameters, $\delta \geq 2\lambda$. In conjunction with the condition in Eq.\eqref{1cond_stab} this imposes a constraint  on the coefficients $\alpha(\phi)$ and $\beta(\phi)$ such that
\begin{equation}
2\lambda\leq\delta<2\lambda+3.  \label{1cond_stab2}
\end{equation}

So far the stability analysis has succeeded to provide a condition for the coefficients. We further need to investigate the evolution of the deceleration parameter $q$ with time for the system obeying Eq.\eqref{1cond_stab} to check if the redshift value at the time of the universe switching from a deceleration phase to an acceleration phase agrees with observations. The redshift $z$ is defined by $z = \frac{a_0}{a(t)}-1$ with $a_0$ being the current value of the scale factor. Since we have defined $N = \ln(a/a_0)$, we can write the redshift in terms of the variable $N$ as
\begin{equation}
z = e^{-N}-1. \label{redshift}
\end{equation}

From the plot of q vs. N we can find out the value of $N$ for which the deceleration parameter $q$ is zero and use that value of N in \eqref{redshift} to obtain the corresponding redshift value. The deceleration parameter can be written in terms of variables $x$ and $y$ as in \eqref{q_exp}.
\begin{equation}
q = -\frac{\dot{H}+H^2}{H^2} = \frac{3}{2}(x+y)+\frac{1}{2} \label{q_exp}.
\end{equation}

\begin{figure*}
	\centering
	\includegraphics[width=0.8\linewidth]{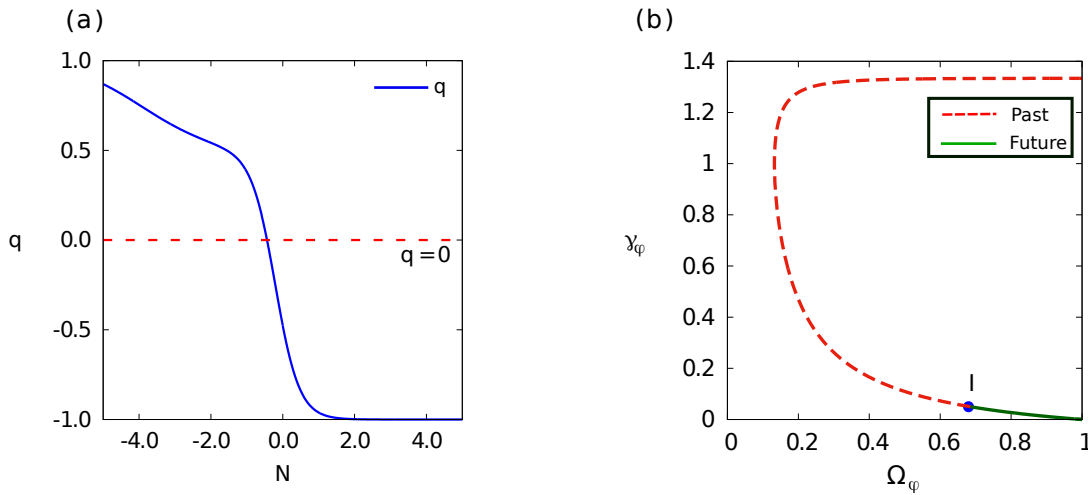}
	\caption{(a) Evolution of the deceleration parameter $q$ with time for $\lambda = 2$, $\delta = 4$. (b) Corresponding $\gp$ vs $\op$ plot. The dashed line shows the backward trajectory and the forward trajectory is shown by the solid line which leads to a scalar field dominated universe.}
	\label{fig:qvsNcase1}
\end{figure*}

\autoref{fig:qvsNcase1}(a) indicates that indeed the universe has changed from a decelerated phase to an accelerated phase of expansion at $z\approx0.6$, close to the observationally estimated value of $z_{obs}\approx 0.5$ \cite{frieman}. The figure also shows that with only the kinetic energy terms in the Lagrangian, the universe will continue  its accelerated expansion phase in the future. \autoref{fig:qvsNcase1}(b) shows the plot of $\gp$ vs $\op$ for $\delta = 2\lambda = 4$. The point I denotes the initial condition given by the current observational values. From the figure, the system evolves to the stable fixed point C with $\gp = 0$ and $\op = 1$, implying that the universe will be completely scalar field dominated.

\subsection*{Nonzero potential ($V(\phi)\neq0$)}
Now we focus our attention to the case where the potential is nonzero. This is a 3-dimensional problem since $b\neq0$. The fixed points of this 3D system and the corresponding eigenvalues are given in \autoref{tab:tab2}. Out of the five fixed points, points C, D and E are the fixed points of the 2D system as well. D and E are unstable as before. The eigenvalues in \autoref{tab:tab2} show that for a stable fixed point to exist at all, the parameter $\sigma= \frac{d(\ln V)}{dN}$ has to be greater than $-3$. Otherwise one of the eigenvalues $(3+\sigma)$ becomes positive and hence the fixed point becomes unstable. So stability condition yields
\begin{equation} \label{sta_2}
\sigma >-3.
\end{equation}
With the condition \eqref{sta_2} both points C and B are stable for $2\lambda>\delta+\sigma$ and depending upon the initial conditions, the trajectories will evolve to either point B or C. Point B is a stable fixed point for $2\lambda<\delta+\sigma$.

\begin{table*}
	\caption{\textbf{Fixed points for the 3-dimensional system with non-zero potential}}\label{tab:tab2}
	\begin{tabular}{D|c|c|D|D|D|c|c}
		\hline Point & x &y & b & $\Omega_\phi$ & $\gamma_\phi$ & Eigenvalues & Stability Condition \\
		\hline \hline A. & 0 & $-\frac{\sigma}{12}$ & $\frac{4+\sigma}{4}$ & 1 & $-\frac{\sigma}{3}$ & \{$\frac{1}{2}(-\delta + 2 \lambda - \sigma)$, $-(3+\sigma)$, $-(4 + \sigma)$\} \TS\B & $2\lambda<\delta+\sigma$ \\ \hline
		B. & $-\frac{\sigma}{6}$ & 0 & $\frac{6+\sigma}{6}$ &  1 & -$\frac{\sigma}{3}$ & \{$(\delta - 2 \lambda + \sigma)$, $-(3+\sigma)$, $-(6 + \sigma)$\} \TS\B & $2\lambda>\delta+\sigma$ \\ \hline
		C. & $\frac{\delta-2\lambda-4}{2}$ & $\frac{6-\delta+2\lambda}{6}$ & 0 & 1 & $\frac{\delta - 2\lambda}{3}$ & \{$\frac{(2\lambda-\delta+6)(2\lambda-\delta+4)}{-2\lambda+\delta-8}$, $(\delta-2\lambda-3)$, $(\delta-2\lambda+\sigma)$\} \TS\B & $2\lambda>\delta+\sigma$ \\ \hline
		D. & 0 & $\frac{1}{3}$ & 0 & 1 & $\frac{4}{3}$ & \{$\frac{1}{2}(4-\delta + 2\lambda$), 1, $(4+\sigma)$\} \TS\B & unstable. \\ \hline
		E. & 1 & 0 & 0 & 1 & 1 & \{3, ($-6+\delta-2\lambda$), $(6+\sigma)$\} \TS\B  & unstable. \\ \hline
	\end{tabular} 
\end{table*}

To determine the initial conditions for the dynamical systems variables $x$, $y$, and $b$, we consider the observables $\op$ and $\gp$ which in this case become,
\begin{subequations}\label{case2obs}
	\begin{align}
	\Omega_\phi &= x + 3y +b, \label{case2op}\\
	\gamma_\phi &= \frac{2x+4y}{x + 3y + b}. \label{case2gp}
	\end{align}
\end{subequations}

\begin{figure}
\centering
\includegraphics[width=1.0\linewidth]{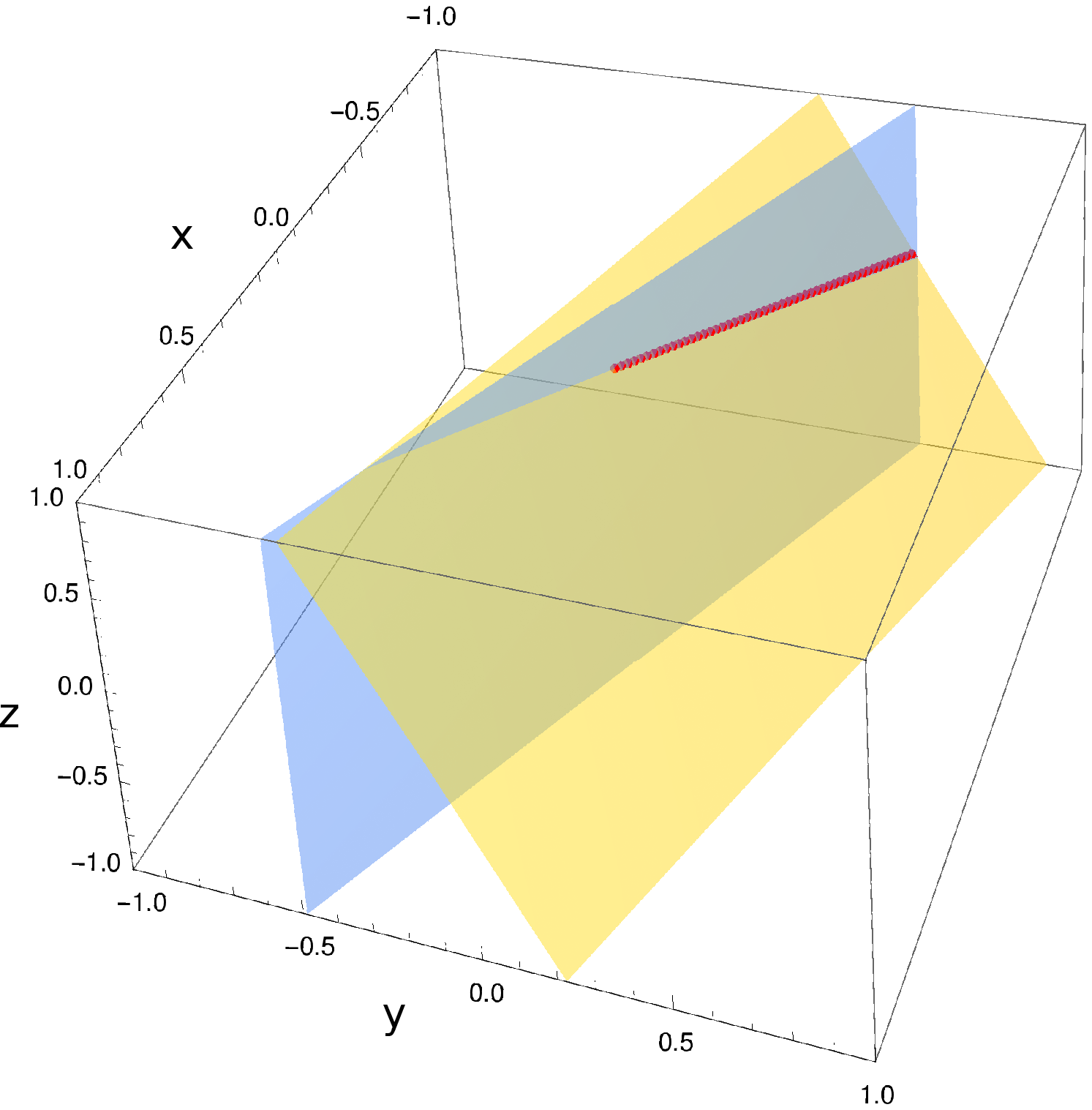}
\caption{The blue and yellow planes show the two surfaces $x+3y+b=0.68$ and $2x+4y=0.034$. The red line shows the initial condition space with the extra condition $y\geq 0$.}
\label{fig:init_3D}
\end{figure}

Using \eqref{case2obs} we can find the fixed point values of the observables as well. As can be seen from \autoref{tab:tab2}, \raisebox{2pt}{$\gamma_\phi$} $= -\frac{\sigma}{3}$ for both the fixed points. To avoid the phantom field models, we must have $\sigma\leq0$, restricting the value of $\sigma$,
\begin{equation}\label{sigma_cond}
-3<\sigma\leq0.
\end{equation}
For this condition, $y$ in point A and $x$ in point B is positive. For point C, $x$ is negative and $y$ is positive as before. For the system to behave analytically in the future, $y$ has to be positive. Otherwise the effective kinetic energy in the Lagrangian \eqref{K_lag} becomes negative, giving rise to a diverging universe (the big rip). So, along with initial conditions \eqref{init} we have $y\geq 0$. But the system is not solvable for $x, y$ and $b$ because we have two equations with three unknowns. So instead of having a particular initial condition, we have a set of initial conditions (\autoref{fig:init_3D}).

Initial conditions with positive x will evolve to point B and initial conditions with negative x will advance to point C given that the stability condition $2\lambda>\delta+\sigma$ is satisfied. Hence the set of initial conditions can be divided into two subsets $x>0$ and $x<0$ each evolving to separate fixed points. It is evident from \eqref{K_lag} and the definition of the variables, that for $y=0$, the Lagrangian \eqref{K_lag} represents quintessence. Initial conditions with $x>0$ ends with quintessence as the ultimate fate. 

\begin{figure*}
	\centering
	\includegraphics[width=0.8\linewidth]{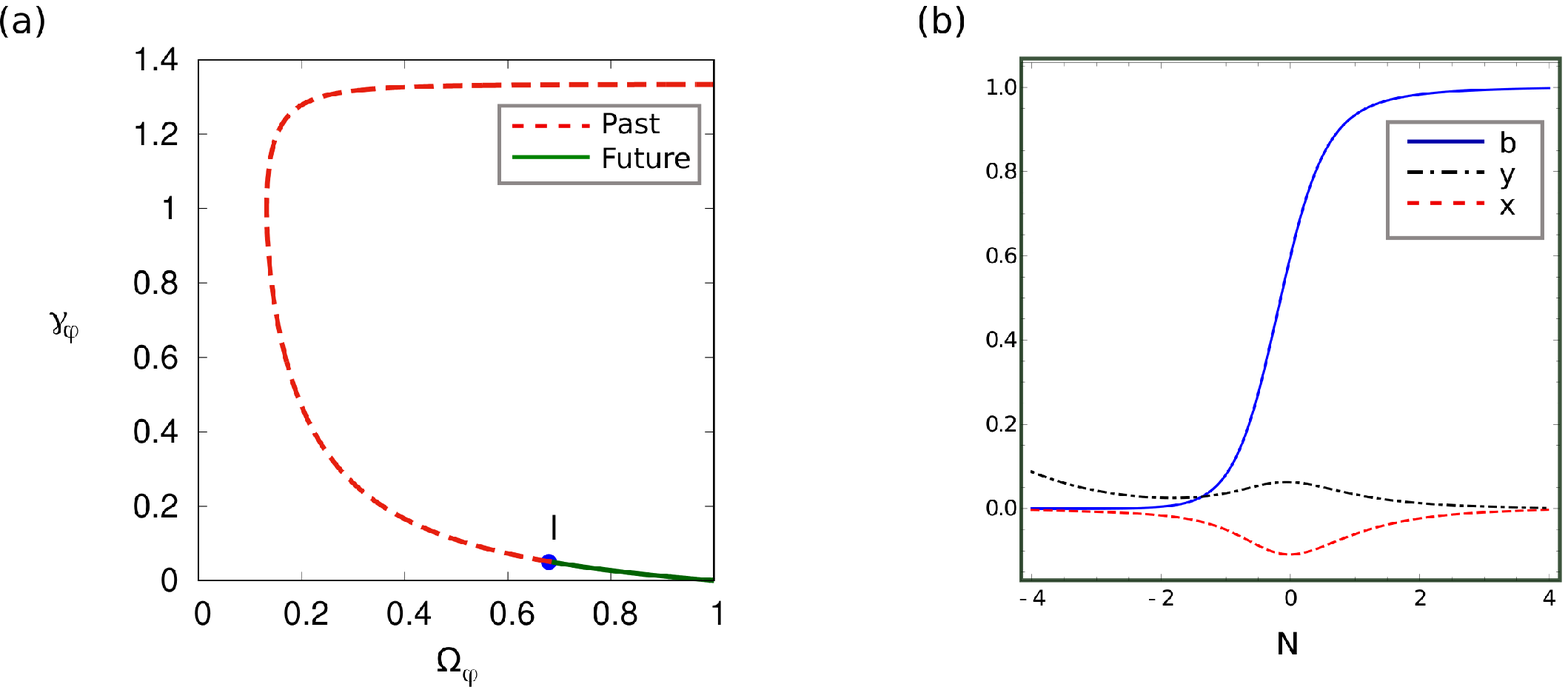}
	\caption{(a) $\gamma_\phi$ vs. $\Omega_\phi$ plot for $\sigma = 0$ and $2\lambda<\delta+\sigma$. The point I denotes the current observation. Dashed line represents the past and solid line represents future trajectory (b) Corresponding time evolution of $b$, $x$, and $y$.}
	\label{fig:gpvsopcase2}
\end{figure*}

\autoref{fig:gpvsopcase2}(a) shows the plot of \raisebox{2pt}{$\gamma_\phi$} vs $\Omega_\phi$ for a limiting value of $\sigma = 0$ and $2\lambda<\delta$. It looks qualitatively similar to \autoref{fig:qvsNcase1}(b). Both of them start from an unstable fixed point $(1,\frac{4}{3})$ and evolves to a stable point $(1,0)$. Still they are fundamentally different. In the first case, the acceleration was driven by the non-canonical kinetic energy term in contrast to the potential driving the acceleration in the second case. This is shown explicitly in the time series plot \autoref{fig:gpvsopcase2}(b). The linear and quadratic kinetic energy terms ($x$ and $y$ respectively) falls off to zero as time increases whereas the potential energy term approaches the maximum value $b=1$, indicating the acceleration of the universe is driven by the potential as in quintessence.

\begin{figure*}
	\includegraphics[width=0.95\linewidth]{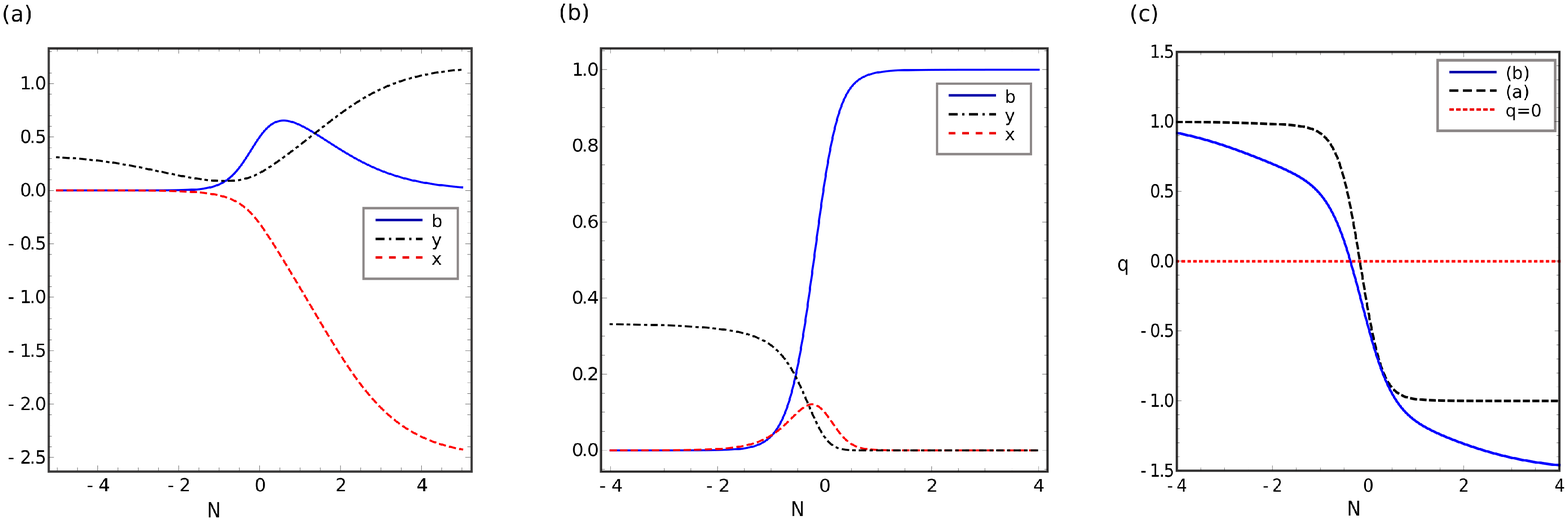}
    \caption{(a) and (b) Time series plots for trajectories evolving to point C and point B respectively. (c) $q$ vs. $N$ plot for the two distinct cases shown in (a) and (b). The black dashed line shows the quintessence like fate and the blue solid line shows the pure K-essence fate. All plots are obtained for $\lambda=1.0$, $\delta=1.0$ and $\sigma=0.0$.}\label{fig:3dq}
\end{figure*}

For $2\lambda>\delta+\sigma$, the universe will either end up with a quintessence scenario or with a pure k-essence case depending upon the choice of initial conditions. \autoref{fig:3dq} (a) shows the timeseries evolution of the variables $x, y$ and $b$ for an initial condition with $x<0$ and we see the potential gradually becomes zero giving rise to a purely kinetic energy dominated universe. \autoref{fig:3dq} (b) on the other hand shows the kinetic energy terms going to zero and the universe ends up with a quintessence like case while starting with $x>0$.

To determine whether the fate of the universe is accelerating or decelerating in this 3-dimensional model, the time evolution of the deceleration parameter $q$ needs to be investigated.  \autoref{fig:3dq}(c) shows the $q$ vs $N$ plots for two distinct cases (case 1 evolving to point B and case 2 evolving to point C). In both cases the accelerated expansion phase continues indefinitely in the future. But quintessence does not yield accurate value for the redshift parameter $z$ when the current phase of acceleration began in the past. The quintessence fate predicts an early redshift value of $z\approx0.2$ while the pure K-essence fate predicts $z\approx0.4$, the latter definitely being closer to the observationally estimated value.

\begin{figure}
\centering
\includegraphics[width=0.7\linewidth]{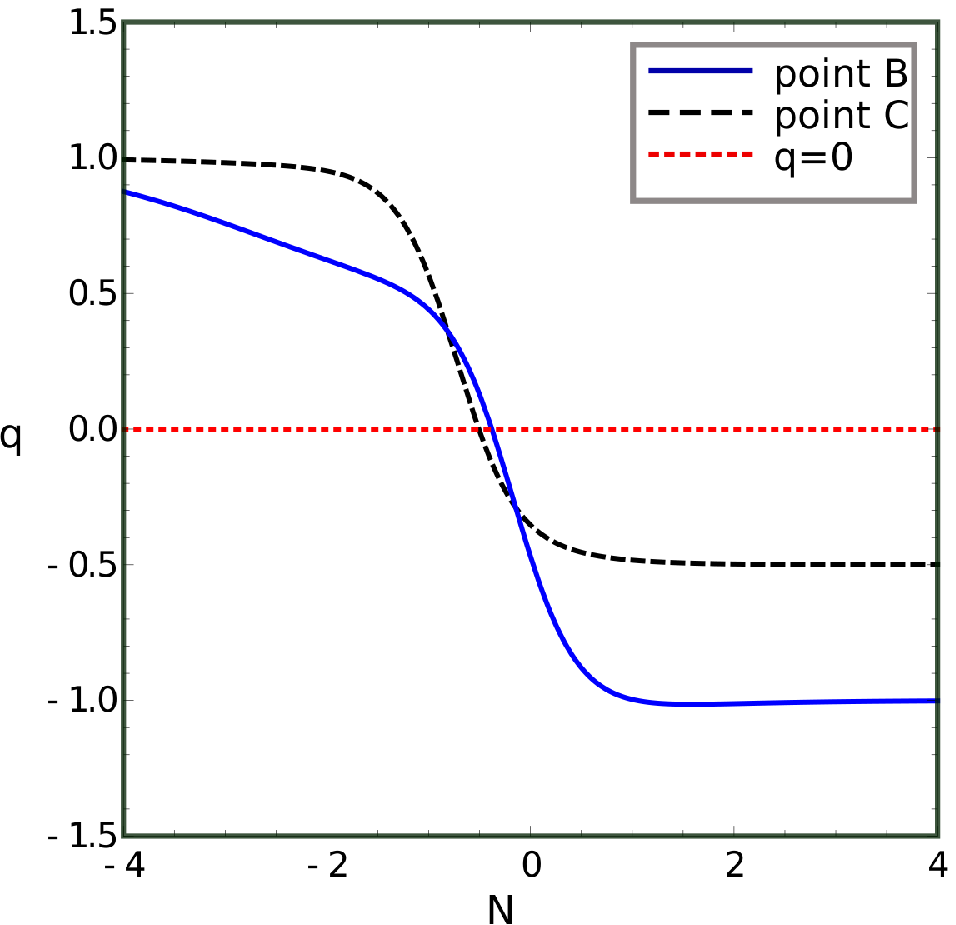}
\caption{$q$ vs $N$ plot for $\sigma\neq0$. The black dashed line shows the time evolution of the deceleration parameter for trajectories evolving to point C (with $b_0<0$) and blue solid line indicates the same for trajectories evolving to point B. }
\label{fig:sigmaneq03d}
\end{figure}

The time evolution of $q$ changes significantly if we change the parameter $\sigma$. For $\sigma\neq0$, evolving the trajectories backward for $x<0$ (point C) and $b>0$, the system hits a singularity. In other words, evolving the pure k-essence like trajectories backward is not possible with a positive potential. With $b_0<0$, however it is possible to get smooth trajectories. On the other hand, the quintessence case is analytic for all $\sigma$ and gives a better prediction of the cosmological redshift values (\autoref{fig:sigmaneq03d}) than the k-essence case for $\sigma\neq 0$ ($z\approx 0.5$ for quintessence like fate, $z\approx0.4$ for pure k-essence like fate).

\section{\label{case2} $\lambda$ and $\delta$ as variables}
In the previous section, $\lambda$ and $\delta$ were treated as two constant parameters of the system. In this section they will be considered as variables. 

\subsection{Pure k-essence}

\begin{table*}
		\caption{\textbf{Fixed points for the 4-dimensional system with zero potential}}\label{tab:tab3}
	\begin{tabular}{D|D|D|C|C|D|D|c|c}
		\hline Point & x & y & $\lambda$ & $\delta$ & $\Omega_\phi$ & $\gamma_\phi$ & eigenvalues \TS\B & Stability\\
		\hline \hline
		A. & 0 & $ \frac{1}{3}$ & 0 & 0 & 1 & $\frac{4}{3}$ & \{2, 1, 1, 1\} \TS\B & Unstable\\ \hline
		B. & 0 & $\frac{1}{3}$ & 0 & $-\frac{4}{4\tau-5}$ & 1 & $\frac{4}{3}$ & \{-1, 1, $\frac{8(\tau-1)}{4\tau-5}$, $\frac{4(\tau-1)}{4\tau-5}$\} \TS\B & Unstable \\ \hline
		C. & 0 & $ \frac{1}{3}$ & $\frac{1}{1-\Gamma}$ & 0 & 1 & $\frac{4}{3}$ & \{-1, 1, 1, $\frac{2\Gamma-3}{\Gamma-1}$\} \TS\B & Unstable\\ \hline
		D. & 0 & $ \frac{1}{3}$ & $-\frac{4(\tau-1)}{(\Gamma-1)(4\tau-5)}$ & $-\frac{4}{4\tau-5}$ & 1 & $\frac{4}{3}$ & \{-1, 1, $-\frac{4(\tau-1)}{4\tau-5}$, $\frac{2(2\Gamma-3)(\tau-1)}{(\Gamma-1)(4\tau-5)}$\} \TS\B & Unstable\\ \hline
		E. & $-2$ & 1 & 0 & 0 & 1 & 0 & \{-3, -3, 0, 0\} \TS\B & Stable \\ \hline
		F. & 1 & 0 & 0 & 0 & 1 & 2 & \{-6, 3, 0, 0\} \TS\B & Unstable \\
		\hline
	\end{tabular} 
\end{table*}

First we consider the pure k-essence, i.e. $b=0$ in \eqref{diff_eq_k}. The system reduces to a 4-dimensional system in this case. The fixed points of the system and the corresponding eigenvalues are given in \autoref{tab:tab3}.
All the fixed points except E are unstable because there is one positive eigenvalue corresponding to each fixed point and the system diverges in that eigendirection. The fixed point E is a nonhyperbolic fixed point. Out of the four eigenvalues, two are zero and the remaining two are negative. One may intuitively conclude that since the remaining nonzero eigenvalues are negative, point E should be stable. To check whether that indeed is the case, numerical simulations are needed. If the trajectories starting close to point E converge back to point E as time evolves then it is a stable fixed point.

\begin{figure*}
	\centering
	\includegraphics[width=1.0\linewidth]{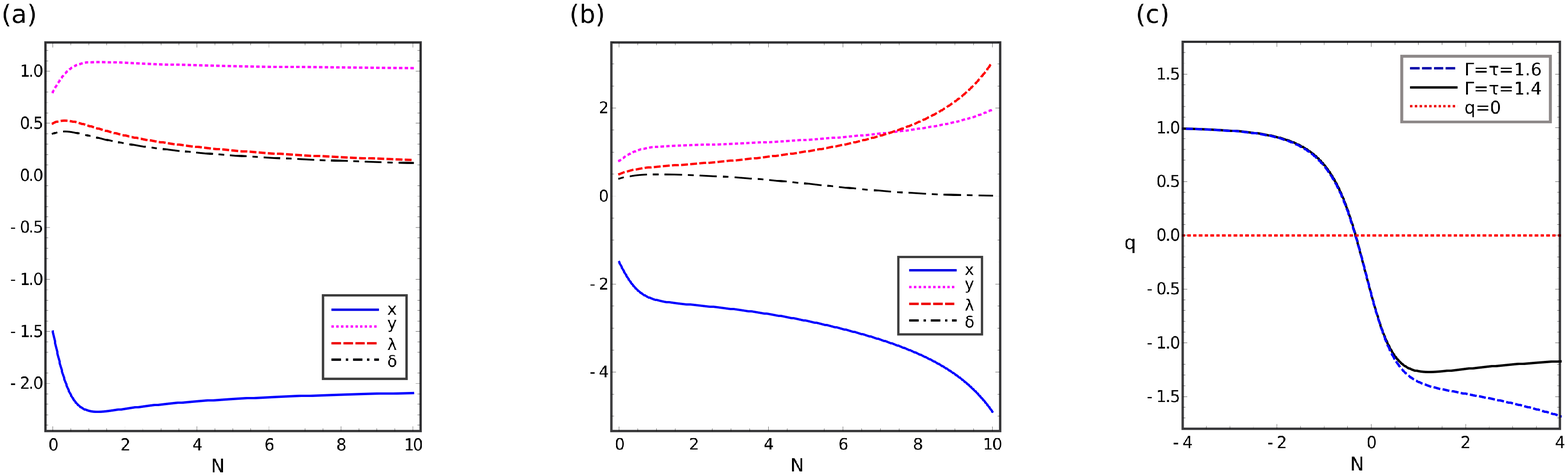}
	\caption{(a) Time evolution of perturbations around fixed point E. $x, y, \lambda, \delta$ for $\Gamma = \tau = 1.4$ (b) Time evolution of perturbations around point E. $\Gamma = \tau = 1.6$. (c) $q$ vs N plot when the system is simulated backward and forward in time, starting with the initial condition \eqref{init} for the two cases in (a) and (b).}
	\label{fig:timeseries4d}
\end{figure*}

Numerical analysis of the system suggests that the point E can either be a stable fixed point or an unstable fixed point depending upon the values of the parameters $\Gamma$ and $\tau$ (\autoref{fig:timeseries4d}(a) and (b)). Starting from the initial conditions \eqref{init} the evolution of the system is exactly same as the 2D system. The fixed point E in \autoref{tab:tab3} can be identified with the fixed point C in \autoref{tab:tab1} with $\delta = \lambda = 0$. This indicates that the $\phi$-dependence of the coefficients disappear as the system approaches the stable fixed point and both $\alpha(\phi)$, and $\beta(\phi)$ become constants. The Lagrangian in this case simplifies to the more familiar form of $\mathcal{L} = F(X)-V(\phi)$ where $F(X)$ is a quadratic polynomial in $X=\dot{\phi}^2/2$.

The change in signature of the deceleration parameter $q$ also happens around the same redshift value as obtained in case 1 with zero potential (\autoref{fig:timeseries4d}(c)). The dynamics only differs if parameters $\Gamma$ and $\tau$ are greater than $1.5$ giving rise to diverging values of $\gamma_{\phi}$ and $\Omega_{\phi}$. By choosing the slope of the co-efficients $\alpha(\phi)$ and $\beta(\phi)$ carefully, these divergences can be avoided.

\subsection*{Nonzero potential ($V(\phi)\neq0$)}
In this section, we consider the most general case with a nonzero potential and $\lambda$, $\delta$ are treated as variables. The system is 5-dimensional as described by the set of equations \eqref{diff_eq_k}. The fixed points and the corresponding eigenvalues for the system are given in \autoref{tab:tab4}. Fixed points F-K are the same fixed points of the 4-D pure k-essence case listed in \autoref{tab:tab3}.
The scalar field energy density parameter $(\Omega_\phi)$ and the effective equation of state parameter ($\gamma_\phi$) are same for the fixed points A-E in \autoref{tab:tab4}, $(1,-\frac{\sigma}{3})$. To avoid a phantom menace, $\sigma$ has to be negative.

\begin{table*}
	\caption{\textbf{Fixed points for the 5-dimensional system with nonzero potential}}\label{tab:tab4}
	\begin{tabular}{D|D|D|c|c|D|c|c}
		\hline Point & $x$ & $y$ & $b$ & $\lambda$ & $\delta$ & Eigenvalues & Stability Condition\\
		\hline \hline
		A. & 0 & $-\frac{\sigma}{12}$ & $\frac{4+\sigma}{4}$ & 0 & 0 & \{$-\frac{\sigma}{2}$, $-\frac{\sigma}{4}$, $-\frac{\sigma}{4}$, $-(\sigma+3)$, $-(\sigma+4)$\}\TS\B & unstable\\ \hline
		B. & 0 & $-\frac{\sigma}{12}$ & $\frac{4+\sigma}{4}$ & $\frac{\sigma}{4(\Gamma-1)}$ & 0 & \{$-\frac{\sigma}{4}$, $\frac{\sigma}{4}$, $-\frac{(2\Gamma-3)\sigma}{4(\Gamma-1)}$, $-(\sigma+3)$, $-(\sigma+4)$\} \TS\B& unstable\\ \hline
		C.  & 0 & $-\frac{\sigma}{12}$ & $\frac{4+\sigma}{4}$ & 0 & $\frac{\sigma}{4\tau-5}$ & \{$\frac{\sigma}{4}$, $-\frac{2(\tau-1)\sigma}{(4\tau-5)}$, $-\frac{(\tau-1)\sigma}{(4\tau-5)}$, $-(\sigma+3)$, $-(\sigma+4)$\} & \specialcell{$\sigma>-3$\\$1<\tau<5/4$}\\ \hline
		D. & 0 & $-\frac{\sigma}{12}$ & $\frac{4+\sigma}{4}$ &  $\frac{\sigma(\tau-1)}{(4\tau-5)(\Gamma-1)}$ &  $\frac{\sigma}{4\tau-5}$ & \{$\frac{\sigma}{4}$,  $\frac{(\tau-1)\sigma}{(4\tau-5)}$,  $-\frac{(2\Gamma-3)(\tau-1)\sigma}{(\Gamma-1)(4\tau-5)}$, $-(\sigma+3)$, $-(\sigma+4)$\} & \specialcell{$\tau>5/4$ or $\tau<1$\\$1<\Gamma<3/2$\\ $\sigma>-3$ } \\ \hline
		E. & $-\frac{\sigma}{6}$ & 0 & $\frac{6+\sigma}{6}$ &  0 & 0 & \{0, 0, $\sigma$, $-(\sigma+6)$, $-(\sigma+3)$\}\TS\B & $\sigma>-3$\\ \hline
		F. & 0 & $ \frac{1}{3}$ & 0 & 0 & 0 & \{2, 1, 1, 1, $4+\sigma$\} \TS\B & Unstable\\ \hline
		G. & 0 & $\frac{1}{3}$ & 0 & 0 & $-\frac{4}{4\tau-5}$ &  \{-1, 1, ($4+\sigma$), $\frac{8(\tau-1)}{4\tau-5}$, $\frac{4(\tau-1)}{4\tau-5}$\} \TS\B & Unstable \\ \hline
		H. & 0 & $ \frac{1}{3}$ & 0 & $\frac{1}{1-\Gamma}$ & 0 & \{-1, 1, 1, $\frac{2\Gamma-3}{\Gamma-1}$, $\sigma + 4$\} \TS\B & Unstable\\ \hline
		I. & 0 & $ \frac{1}{3}$ & 0 & $-\frac{4(\tau-1)}{(\Gamma-1)(4\tau-5)}$ & $-\frac{4}{4\tau-5}$ & \{-1, 1, $\sigma + 4$, $-\frac{4(\tau-1)}{4\tau-5}$, $\frac{2(2\Gamma-3)(\tau-1)}{(\Gamma-1)(4\tau-5)}$\} \TS\B & Unstable\\ \hline
		J. & $-2$ & 1 & 0 & 0 & 0 & \{-3, -3, $\sigma$, 0, 0 \} \TS\B & $\sigma<0$ \\ \hline
		K. & 1 & 0 & 0 & 0 & 0 & \{-6, 3, 0, 0, $(6+\sigma)$\} \TS\B & Unstable \\
		\hline
	\end{tabular}
	\end{table*} 
	
From \autoref{tab:tab4}, point A and B will never be a stable fixed point since at least one of the eigenvalues is positive in both cases. Depending upon conditions on $\Gamma$ and $\tau$ point C and D can be stable or unstable. The choice for valid forms of $\alpha(\phi)$ and $\beta(\phi)$ will be severely restricted by these conditions. Point E on the other hand gives a stable node without any further restriction on $\Gamma$ and $\tau$. Hence it allows more generic forms of $\alpha(\phi)$ and $\beta(\phi)$.  However, the stability of point E is still dependent on the steepness of the potential ($\sigma > -3$). Out of the other fixed points indicating the pure k-essence case, only point J is a stable fixed point (non hyperbolic fixed point) for $\Gamma$, $\tau < 1.5$ and $\sigma<0$. 

\begin{figure*}
	\centering
	\includegraphics[width=0.8\linewidth]{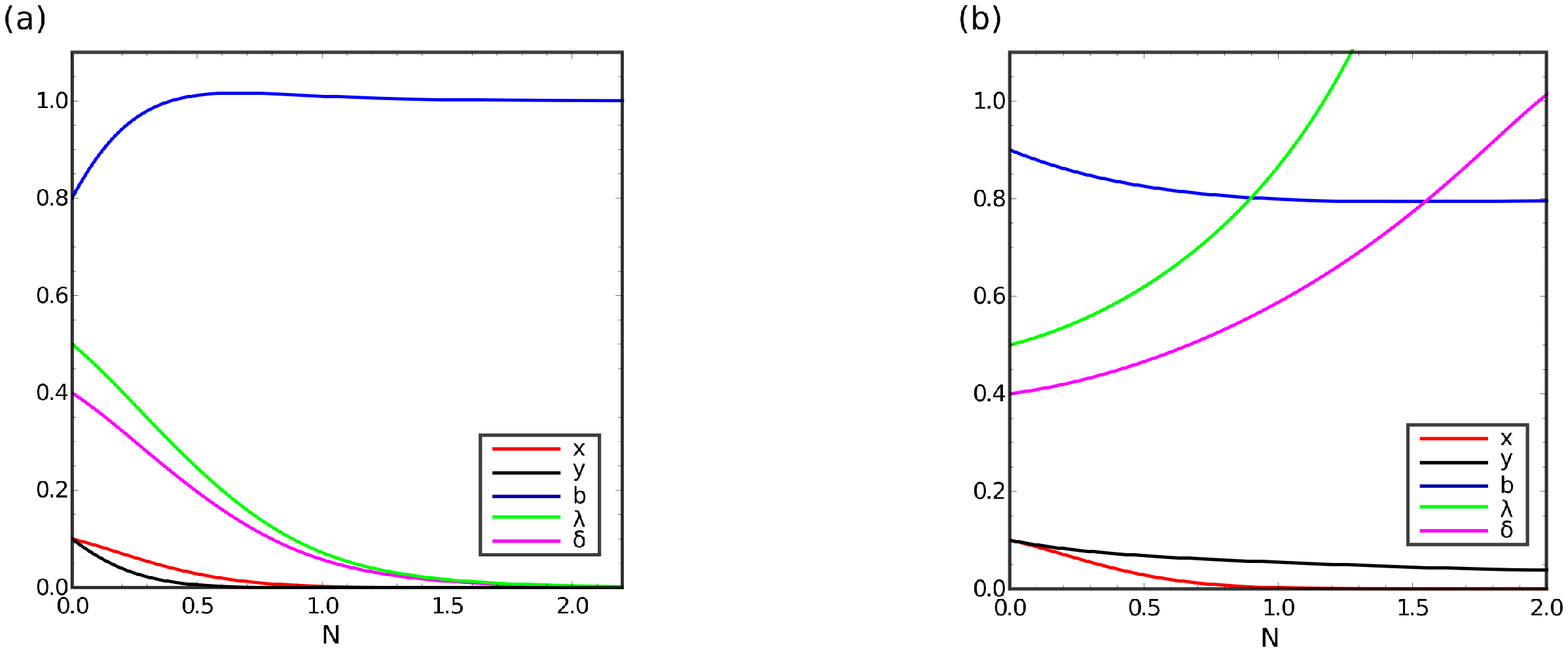}
	\caption{(a) Time series evolution of five variables for $\Gamma = \tau = 1.0$. All the variables converge to the fixed point E. (b) For $\Gamma = \tau = 2.0$ and $\sigma \neq 0$, $x, y, b$ converges to their respective values corresponding to fixed point E. However, $\lambda$ and $\delta$ show diverging trajectories.}
	\label{fig:case5time}
\end{figure*}

The fixed point E is actually a nonhyperbolic fixed point.  So eigenvalue analysis cannot rightly predict the stability of the fixed point. To check the stability, the 5 dimensional system is simulated numerically with the initial conditions taken in a neighborhood of the fixed point. The $x$, $y$, $b$ values converge to the fixed point values in \autoref{tab:tab4}, but depending upon $\Gamma$, $\tau$ and $\sigma$, either $\lambda$ or $\delta$ or both diverge (\autoref{fig:case5time}). Since the observables $\gamma_\phi$ and $\Omega_\phi$ are independent of the variables $\lambda$ and $\delta$, we do not see any observable effect of the divergences in these two variables on the system.

\begin{figure}[ht]
	\centering
	\includegraphics[width=0.7\linewidth]{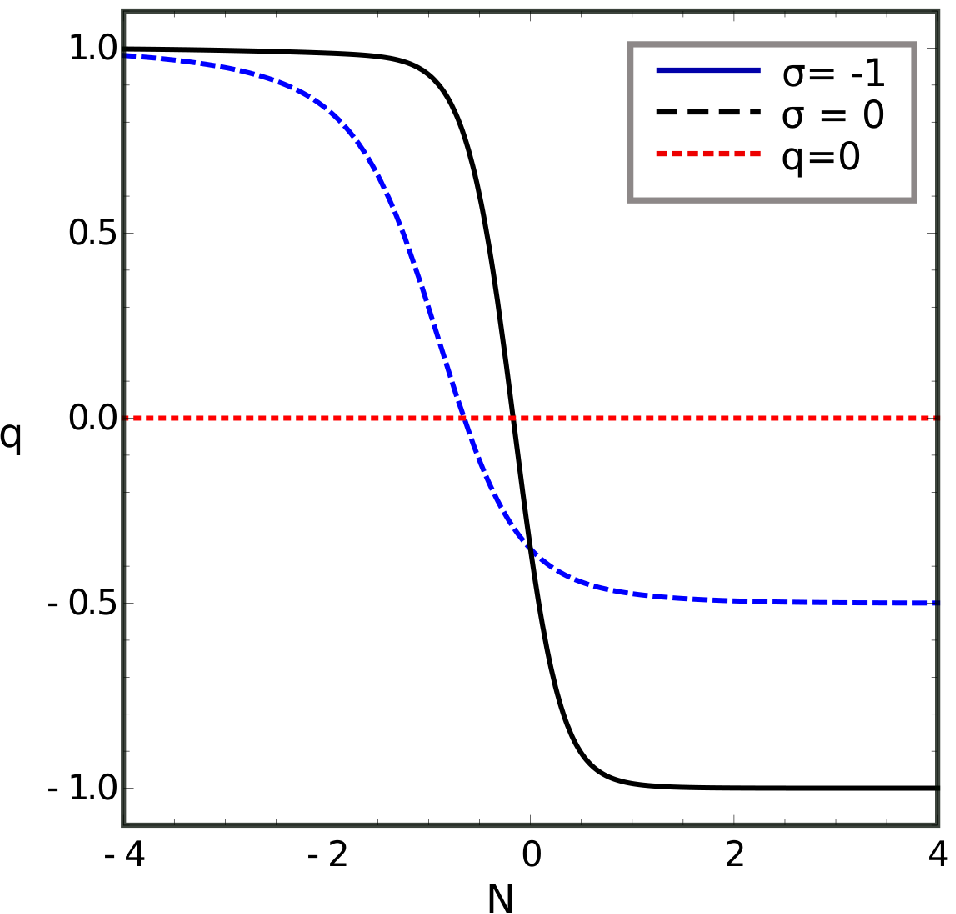}
	\caption{q vs N plot for two different parameter values $\sigma = 0$ (black line) and $\sigma = -1$ (blue line).}
	\label{fig:qvsNcase4}
\end{figure}

As in the 3-D system with nonzero potential, unique solutions for initial conditions of $x$, $y$ and $b$ cannot be obtained from the initial conditions \eqref{case2obs} since we have only two equations with three variables ($x$, $y$, and $b$). Instead we have a set of initial conditions as shown in \autoref{fig:init_3D}. Similar to the 3D case, initial conditions with $x>0$ and $y>0$ evolves to point E and initial conditions with $x<0$ and $y>0$ advances to point J (pure k-essence). For the fixed point E, $y=0$, i.e. the non-canonical kinetic energy term in the Lagrangian becomes zero, describing the quintessence case. This demonstrates that even if we start with a k-essence model, the universe can end up with a quintessence scenario in the long run. Though there can be other fixed points, all of them imposes further constraints on the allowed forms of the coefficients $\alpha(\phi)$ and $\beta(\phi)$. Also, trajectories starting from the initial condition space cannot evolve to the fixed points C and D. Point E on the other hand provides a more generic choice of the coefficients. And it allows us consistent trajectories with varying $\sigma$ without encountering any singularities in the near past, in contrast with the k-essence case. We can say that the quintessence case is the most generic choice for the ultimate fate of the universe for a k-essence model. In the quintessence case, the final value of $\lambda$ and $\delta$ are zero, indicating the coefficients $\alpha$ and $\beta$ become constants near the fixed point.
To check for the shift in the signature of the deceleration parameter $q$, the system is evolved with time for two different values of the parameter $\sigma$. In both the cases the universe ends up in an accelerated expansion phase (\autoref{fig:qvsNcase4}). The prediction for the redshift value is closer ($z\approx0.6$) to the observation value for $\sigma\neq0$.

\section{Discussion}
An investigation of the k-essence model (with quadratic terms in kinetic energy) from the perspective of stability of the model has been carried out in the present work. The discussion divided into two main sections: (a) $\lambda, \delta$ treated as parameters of the system and (b) $\lambda, \delta$ treated as variables. Both sections are divided into two subsections with zero and non-zero potential, i.e. pure k-essence and k-essence with a scalar field dependent potential. 
So four broad classes have been discussed. The investigation is quite exhaustive for the Lagrangian allowing upto the quadratic powers of the kinetic energy.\\

From the stability analysis, one can see that depending on the choices of the parameters, all the four classes can lead to a stable model of the universe which settles in an accelerated phase of expansion. The fixed points are all tabulated in tables I to IV. We now summarize the properties of the stable fixed points only. Fixed point C in \autoref{tab:tab1} correctly describes the kinematics of the universe, but the trajectory may diverge in the reverse direction. The fixed point A in \autoref{tab:tab2} is well-behaved for $\sigma =0$ which means a constant potential, essentially a cosmological constant. So there is hardly any advantage of the choice of an exotic kinetic term. Fixed point B in \autoref{tab:tab2} is favorable in terms of stability and the kinematics of the universe. But this essentially evolves to a standard quintessence model. Point C in the same table again leads essentially to a cosmological constant as it works well only for $\sigma =0$. Point E in \autoref{tab:tab3} is identical with C in table \autoref{tab:tab1}, which is already discussed. Points C and D in \autoref{tab:tab4} suffer from the fact that the trajectories from the viable set of initial conditions do not evolve to this fixed point. Point J again leads effectively to a cosmological constant. Fixed point E in table \autoref{tab:tab4} seems to be working well in terms of stability and kinematics.  It is easy to see that this evolves to a quintessence scenario as well. \\

Thus, albeit the work is done for a k-essence containing upto a second order in the kinetic energy, this investigation leads to an important indication. Stability criteria and kinematic requirements taken together, only those k-essence models are viable which actually evolve to a quintessence model. So the noncanonical kinetic part is not really favored. \\

Acknowledgement: AC wishes to thank KVPY, funded by Department of Science and Technology, India, for financial support.

\vfill
\end{document}